\documentclass[aps,twocolumn, amsmath, amssymb, a4paper, 10pt, superscriptaddress]{revtex4-2}

\usepackage{graphicx}
\usepackage[utf8]{inputenc}
\usepackage{bm}
\usepackage{xcolor}
\usepackage{physics}
\usepackage[colorlinks, citecolor=blue, linkcolor=blue]{hyperref}
\graphicspath{{Figures/}}
\usepackage{siunitx}
\usepackage{xspace}
\usepackage{nicefrac}

\newcommand{\BFO}{BiFeO\textsubscript{3}\xspace}

\begin{document}

\title{Imaging topological defects in a non-collinear antiferromagnet}

\author{Aurore Finco}
\author{Angela Haykal}
\affiliation{Laboratoire Charles Coulomb, Université de Montpellier and CNRS, 34095 Montpellier, France}
\author{Stéphane Fusil}
\affiliation{Unité Mixte de Physique, CNRS, Thales, Université Paris-Saclay, 91767 Palaiseau, France}
\author{Pawan Kumar}
\affiliation{Laboratoire Charles Coulomb, Université de Montpellier and CNRS, 34095 Montpellier, France}
\author{Pauline Dufour}
\affiliation{Unité Mixte de Physique, CNRS, Thales, Université Paris-Saclay, 91767 Palaiseau, France}
\author{Anne Forget}
\author{Dorothée Colson}
\author{Jean-Yves Chauleau}
\author{Michel Viret}
\affiliation{SPEC, CEA, CNRS, Université Paris-Saclay, 91191 Gif sur Yvette, France}
\author{Nicolas Jaouen}
\affiliation{Synchrotron SOLEIL, 91192 Gif-sur-Yvette, France}
\author{Vincent Garcia}
\affiliation{Unité Mixte de Physique, CNRS, Thales, Université Paris-Saclay, 91767 Palaiseau, France}
\author{Vincent Jacques}
\affiliation{Laboratoire Charles Coulomb, Université de Montpellier and CNRS, 34095 Montpellier, France}


\begin{abstract}

We report on the formation of topological defects emerging from the cycloidal antiferromagnetic order at the surface of bulk \BFO crystals. Combining reciprocal and real-space magnetic imaging techniques, we first observe, in a single ferroelectric domain, the coexistence of antiferromagnetic domains in which the antiferromagnetic cycloid propagates along different wavevectors. We then show that the direction of these wavevectors is not strictly locked to the preferred crystallographic axes as continuous rotations bridge different wavevectors. At the junctions between the magnetic domains, we observe topological line defects identical to those found in a broad variety of lamellar physical systems with rotational symmetries. Our work establishes the presence of these magnetic objects at room temperature in the multiferroic antiferromagnet \BFO, offering new possibilities for their use in spintronics.  

\end{abstract}
\date{\today}

\maketitle


When symmetry-breaking phase transitions occur, they induce the formation of topological defects~\cite{zurekCosmologicalExperimentsCondensed1996}, which are isolated regions of lower dimensionality corresponding to a singularity of the order parameter~\cite{merminTopologicalTheoryDefects1979}. From topology and symmetry considerations, these defects can be identified and classified in various physical systems, from cosmological to nanometric length scales. In lamellar structures with rotational symmetries, such singular lines appear as dislocations and disclinations which have been extensively studied~\cite{deGennesLiquidCrystals}, even before the topological approach to their description was proposed. Illustrating the universality of these patterns~\cite{seul1995domain}, similar defects can be found in a broad variety of modulated physical systems including cholesteric liquid crystals~\cite{Bouligand}, ferromagnetic garnets~\cite{seul1992evolution} or copolymers~\cite{murphy2015automated}. Recently, they were also identified in the helical state of the ferromagnetic B20 material FeGe~\cite{schoenherrTopologicalDomainWalls2018a}, providing -- besides skyrmions~\cite{yuRoomtemperatureFormationSkyrmion2011} -- additional topological magnetic textures for future use in spintronics.

In this work, we demonstrate that these topological defects are present in another type of chiral magnetic material, the multiferroic \BFO, which exhibits a lamellar pattern originating from an antiferromagnetic cycloidal order~\cite{sosnowskaSpiralMagneticOrdering1982} rather than a ferromagnetic helix. We start with a detailed depiction of the magnetic texture of bulk \BFO, and we discuss the origin of the generated stray field allowing us to achieve real-space imaging of the antiferromagnetic state with scanning-NV magnetometry. Then, we use this technique to investigate quantitatively the magnetic order and show the coexistence of three magnetic rotational domains within a single ferroelectric domain. At the junctions between these areas where the cycloids propagate along different directions, we observe either a smooth rotation of the wavevector or the formation of topological defects. 


\begin{figure*}[t]
\includegraphics[width=17.8cm]{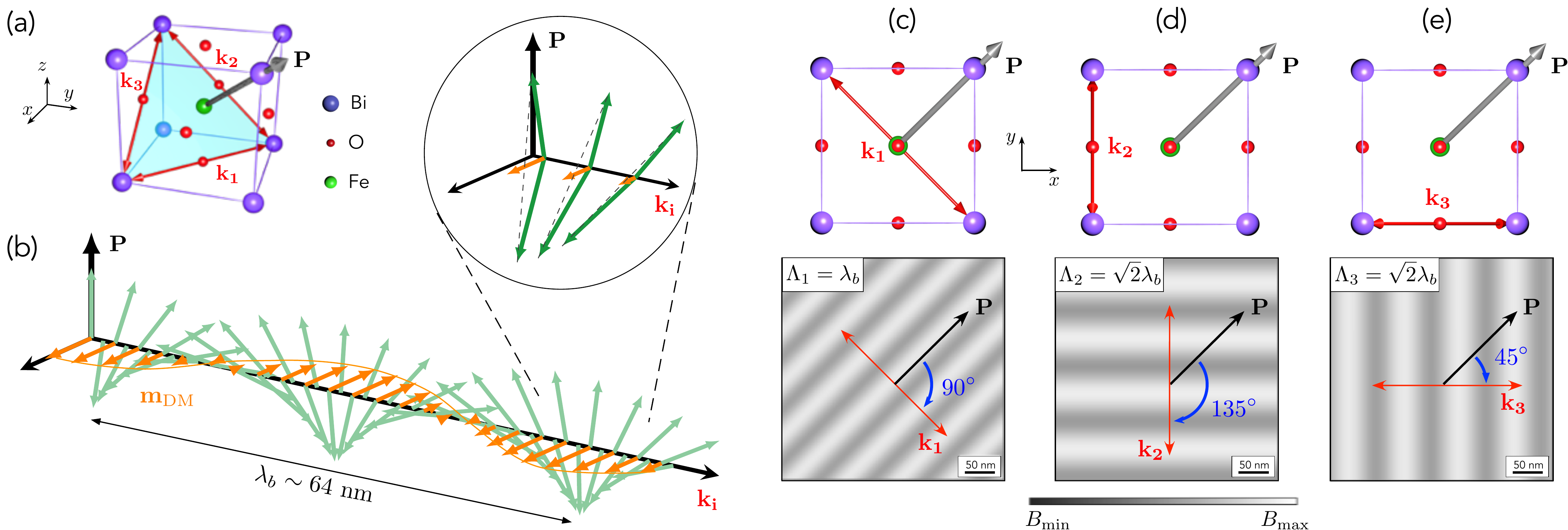}
\caption{
{\bf (a)} Sketch of the pseudo-cubic cell of \BFO showing the three possible propagation directions of the antiferromagnetic cycloid for a given ferroelectric variant. Here $\mathbf{P}$ is pointing along the [111] axis, so that $\mathbf{k_1}\parallel[-110]$, $\mathbf{k_2}\parallel[01-1]$ and $\mathbf{k_3}\parallel[10-1]$. {\bf (b)} Schematic representation of the cycloidal antiferromagnetic order, which propagates with a wavevector $\mathbf{k_i}$ perpendicular to the ferroelectric polarization $\mathbf{P}$. The zoom illustrates the canting of the spins out of the cycloidal plane, leading to a spin density wave (SDW) characterized by the modulated uncompensated moment $\mathbf{m_{\rm DM}}$. {\bf (c-e)} Top panels: Projection of the three possible propagation directions of the cycloid on a (001) surface. Bottom panels: Expected magnetic field maps produced by the SDW above a (001)-oriented \BFO sample, showing a modulated field amplitude whose direction and period is tied to the cycloid wavevector.
}
\label{fig1}
\end{figure*}

At room temperature, bulk \BFO is characterized by a slightly distorted rhombohedral structure, which is commonly described by the perovskite-type pseudo-cubic unit cell shown in Fig.~\ref{fig1}(a)~\cite{moreauFerroelectricBiFeO3Xray1971, kubelStructureFerroelectricFerroelastic1990}. The Bi\textsuperscript{3+} ions are displaced with respect to the FeO\textsubscript{6} octahedra, giving rise to a large spontaneous ferroelectric polarization $\mathbf{P}$ pointing along one of the $\langle 111 \rangle$ directions of the pseudo-cubic cell~\cite{wangEpitaxialBiFeO3Multiferroic2003,lebeugleRoomtemperatureCoexistenceLarge2007}. Besides this ferroelectric behavior, \BFO is also an antiferromagnet with a N\'eel temperature $T_{\rm N}= \SI{643}{\K}$. The superexchange interaction mediated by the Fe-O-Fe bond favors a collinear, G-type antiferromagnetic order. Moreover, the slight shift of the Fe\textsuperscript{3+} ions away from the center of the FeO\textsubscript{6} octahedra induces an additional antisymmetric exchange interaction, often referred to as magnetoelectric interaction, which promotes a non-collinear ordering of the magnetic moments~\cite{katsuraSpinCurrentMagnetoelectric2005}. The competition between the superexchange and the magnetoelectric interactions results in the stabilization of an incommensurate cycloidal rotation of the Fe\textsuperscript{3+} magnetic moments propagating along a direction $\mathbf{k}$ perpendicular to the ferroelectric polarization. In bulk \BFO, the period $\lambda_b$ of the cycloidal antiferromagnetic order is around \SI{64}{\nano\meter}~\cite{lebeugleElectricFieldInducedSpinFlop2008, sosnowskaSpiralMagneticOrdering1982}. Importantly, for a given polarization direction, the rhombohedral symmetry of \BFO allows three equivalent propagation directions of the cycloid, thus forming three rotational magnetic domains~\cite{ramazanogluLocalWeakFerromagnetism2011}. If $\mathbf{P}$ is along the [111] axis, the three propagation vectors $\mathbf{k_1}$, $\mathbf{k_2}$ and $\mathbf{k_3}$ lie in a (111) plane and point along the $[1-10]$, $[0-11]$ or $[10-1]$ directions, as sketched in Fig.~\ref{fig1}(a). This non-collinear antiferromagnetic order, which is referred to as \emph{type I} cycloid~\cite{sandoCraftingMagnonicSpintronic2013, agbeleleStrainMagneticField2017, haykalAntiferromagneticTexturesBiFeO32020,Sando2020}, is the one commonly observed in bulk \BFO~\cite{lebeugleElectricFieldInducedSpinFlop2008, sosnowskaSpiralMagneticOrdering1982, ramazanogluLocalWeakFerromagnetism2011}.

The rotation angle induced by the magnetoelectric coupling is constant from one Fe site to another along the cycloid propagation direction~\cite{katsuraSpinCurrentMagnetoelectric2005}. Each magnetic moment is thus compensated by its nearest antiparallel neighbors, a fact commonly overlooked in many previous descriptions of the cycloid~\cite{lebeugleElectricFieldInducedSpinFlop2008,grossRealspaceImagingNoncollinear2017}. On a scale larger than the inter-atomic distance, an antiferromagnetic cycloid resulting solely from the competition between the superexchange and the magnetolectric interactions is therefore perfectly compensated, and does not produce any static magnetic fields for compensated (001) surfaces. To understand the origin of the stray magnetic field recently measured above \BFO thin films~\cite{haykalAntiferromagneticTexturesBiFeO32020,chauleauElectricAntiferromagneticChiral2020,grossExploringNoncollinearSpin2017}, an additional uncompensation is required. It originates from a Dzyaloshinskii-Moriya (DM) interaction stemming from the antiphase rotation of the oxygen octahedra along the [111] direction~\cite{ramazanogluLocalWeakFerromagnetism2011}. This interaction cants the spins out of the cycloidal plane, leading to an uncompensated magnetic moment $\mathbf{m_{\rm DM}}$ with a periodic magnitude locked to that of the antiferromagnetic cycloid, as sketched in Fig.~\ref{fig1}(b). The resulting spin density wave (SDW) generates a stray magnetic field whose spatial distribution depends on the cycloid propagation vector. In this work, we employ scanning-NV magnetometry to image this magnetic field distribution in real space with nanoscale spatial resolution.

To discriminate between the three possible propagation directions of the cycloid within a single ferroelectric domain, it is convenient to examine their projections on the sample surface. Considering a conventional (001)-oriented \BFO crystal, the $\mathbf{k_1}$ propagation vector lies in the surface plane, making an angle of \ang{90} with the in-plane component of the polarization [see Fig.~\ref{fig1}(c)]. The corresponding SDW produces a stray magnetic field modulated along $\mathbf{k_1}$ with a period $\Lambda_1=\lambda_b\sim \SI{64}{\nano\meter}$. On the other hand, the $\mathbf{k_2}$ and $\mathbf{k_3}$ propagation vectors are lying out of the surface plane, with an in-plane projection  making an angle of \ang{135} and \ang{45} with the in-plane polarization, respectively [Fig.~\ref{fig1}(d,e)]. In both cases, the period of the stray field modulation is equal to the projection of the intrinsic cycloid wavelength on the (001) surface plane, leading to $\Lambda_2=\Lambda_3=\sqrt{2}\,\lambda_b\sim \SI{90}{\nano\meter}$. The period and the orientation of the magnetic field pattern can therefore be used to identify the propagation vector of the cycloid.

\begin{figure}[t]
\includegraphics[width=8.7cm]{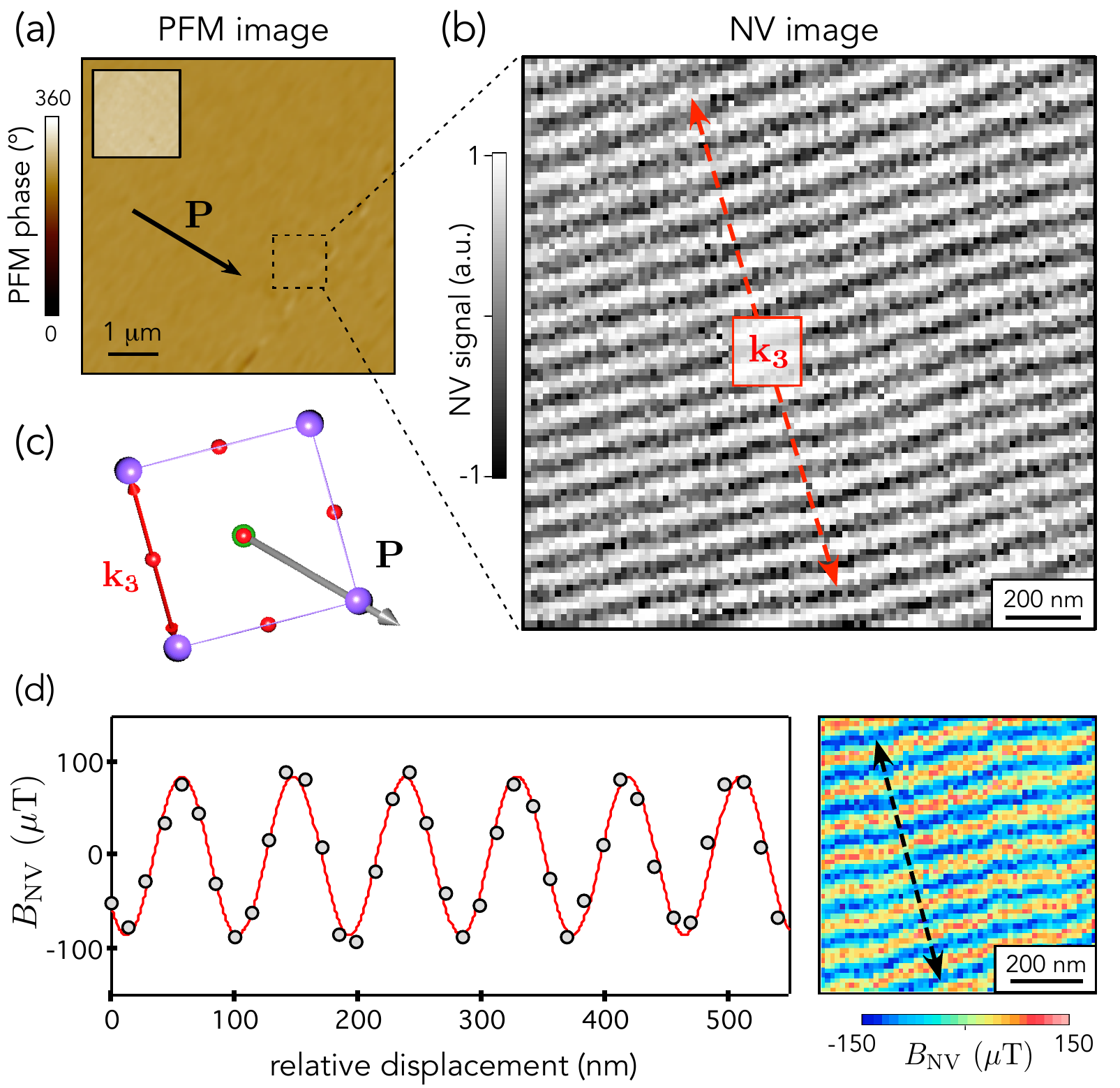}
\caption{{\bf (a)} Out-of-plane PFM phase image recorded above the (001)-oriented bulk \BFO crystal. Inset: corresponding in-plane PFM phase image. The black arrow indicates the projection of the ferroelectric polarization $\mathbf{P}$ on the (001) sample surface. {\bf (b)} Magnetic field image recorded with the scanning-NV magnetometer operated in dual-iso-B imaging mode. The red dashed line indicates the propagation direction of the cycloid along $\mathbf{k_3}$. The characteristic period of the modulation is obtained by fitting line profiles with a sinusoidal function, leading to $\Lambda_3= \SI[multi-part-units=single]{92 \pm 2}{\nano\meter}$. {\bf (c)} Sketch illustrating that the recorded magnetic image is only compatible with a spin cycloid propagating along $\mathbf{k_3}$. 
{\bf (d)} Right panel: fully quantitative magnetic field distribution. Left panel: line profile taken along the cycloid propagation direction. The red solid line is data fitting (see main text) yielding an uncompensated moment $m_{\rm DM}=0.09\pm 0.03 \ \mu_{\rm B}$. 
}
\label{fig2}
\end{figure}

Below, we investigate the antiferromagnetic order in a millimeter-sized (001)-oriented bulk \BFO crystal grown by the Bi\textsubscript{2}O\textsubscript{3}-Fe\textsubscript{2}O\textsubscript{3} flux method~\cite{lebeugleRoomtemperatureCoexistenceLarge2007}. The local ferroelectric properties were first characterized through piezoresponse force microscopy (PFM). As shown in Fig.~\ref{fig2}(a), the \BFO crystal exhibits a single ferroelectric domain, a feature that is commonly observed for bulk crystals grown by the flux method at temperatures much lower than the ferroelectric Curie temperature ($T_{\rm C}\sim \SI{1100}{\K}$)~\cite{lebeugleElectricFieldInducedSpinFlop2008}. Combining in-plane and out-of-plane PFM measurements, the direction of the ferroelectric polarization $\mathbf{P}$ was unambiguously identified [see black arrow in Fig.~\ref{fig2}(a)]. The corresponding antiferromagnetic texture was imaged in real space with a scanning-NV magnetometer operating under ambient conditions. A commercial diamond tip hosting a single NV defect at its apex (Qnami, Quantilever MX) was integrated into an atomic force microscope and scanned in close proximity of the bulk \BFO crystal surface. At each point of the scan, a quantitative magnetic field measurement was obtained by monitoring the Zeeman shift of the NV defect electron spin sublevels through optical detection of the magnetic resonance~\cite{rondinMagnetometryNitrogenvacancyDefects2014}. The NV-to-sample distance, which sets the spatial resolution of the magnetic microscope, was inferred through an independent calibration procedure~\cite{hingantMeasuringMagneticMoment2015}, leading to $d_{\rm NV}=\SI[multi-part-units=single]{60\pm1}{\nano\meter}$~\cite{suppl}.

The scanning-NV magnetometer was first operated in dual-iso-B imaging mode to obtain a fast characterization of the magnetic field distribution produced by the SDW bounded to the cycloidal antiferromagnetic order. A typical dual-iso-B image recorded above a single ferroelectric domain of the bulk \BFO crystal is shown in Fig.~\ref{fig2}(b). Here the magnetometer signal is modulated along a direction that makes an angle of about \ang{45} with the in-plane component of the polarization. This observation indicates an out-of-plane propagation of the cycloid along $\mathbf{k_3}$ [Fig.~\ref{fig2}(c)], which is further supported by the measured surface period of the modulation $\Lambda_3=\SI[multi-part-units=single]{90\pm 2}{\nano\meter}$. In this micron-sized area of the sample, we thus observe a single rotational magnetic domain. This experiment constitutes the first real-space observation of the non-collinear antiferromagnetic order in a bulk \BFO crystal.

To infer an estimate of the uncompensated magnetic moment $m_{\rm DM}$ linked to the spin cycloid, a fully quantitative magnetic field image was recorded~\cite{rondinMagnetometryNitrogenvacancyDefects2014}. The resulting map shows a magnetic field modulation with a typical amplitude in the range of $\SI{100}{\micro\tesla}$ [Fig.~\ref{fig2}(d)]. A quantitative analysis of this magnetic field distribution was performed by modeling the SDW at a position $\mathbf{r}$ in the \BFO crystal as
\begin{equation}
\mathbf{m_{\rm DM}}(\mathbf{r})=m_{\rm DM}\cos(\mathbf{k_3} \cdot \mathbf{r})(\mathbf{e_{k_3}}\times \mathbf{e_{P}}) \ ,
\end{equation}
where $\|\mathbf{k_3}\|=2\pi/\lambda_b$, and $\{\mathbf{e_{k_3}},\mathbf{e_{P}}\}$ are orthogonal unit vectors oriented along the cycloid propagation direction and the polarization [see Fig.~\ref{fig1}(b)]. An analytical calculation of the stray field produced by such a spatial distribution of magnetic moments is given in Supplemental Material~\cite{suppl}. The resulting formula was used to fit line profiles across the recorded stray field map with $m_{\rm DM}$ as fitting parameter [Fig.~\ref{fig2}(d)]. This analysis leads to $m_{\rm DM}=0.09\pm 0.03 \ \mu_{\rm B}$, a value in good agreement with the one inferred through polarized neutron scattering experiments in bulk \BFO crystals~\cite{ramazanogluLocalWeakFerromagnetism2011}. It corresponds to a periodic wiggling of the cycloidal plane with a maximal canting angle of about $\ang{1}$ [Fig.~\ref{fig1}(b)]. We note that the uncertainty of the measurement is here mainly due the imperfect knowledge of the flying distance of the scanning-NV sensor~\cite{grossRealspaceImagingNoncollinear2017} (see Supplemental Material~\cite{suppl}). 

\begin{figure}[t]
\includegraphics[width=8.6cm]{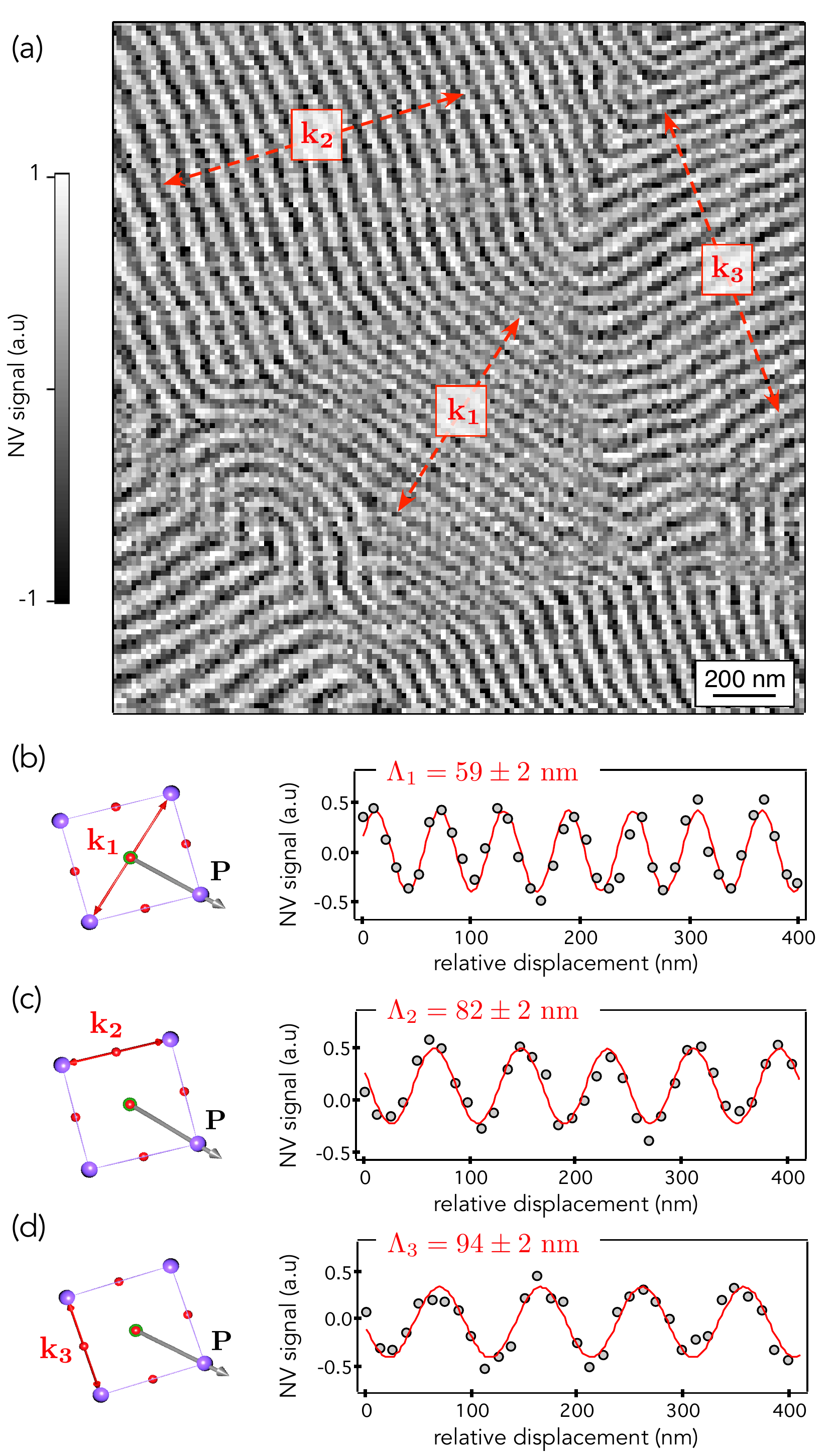}
\caption{(a) NV magnetometry image recorded in dual-iso-B mode showing the coexistence of the three propagation directions of the antiferromagnetic cycloid (red dashed arrows) in a single ferroelectric domain of the bulk \BFO crystal. (b-d) The left panels show sketches of the three projected propagation directions on the (001) surface with respect to the in-plane component of the polarization $\mathbf{P}$. The right panels show line profiles across the three propagation directions. Solid lines are data fitting with a cosine function and the corresponding periods resulting from the fits are indicated.}
\label{fig3}
\end{figure}

With this set of results, it is tempting to conclude that the antiferromagnetic order in our bulk \BFO crystal can be simply described by a single spin cycloid variant, as reported previously using neutron scattering experiments~\cite{lebeugleElectricFieldInducedSpinFlop2008}. However, magnetic measurements performed on different regions of the crystal reveal more complex configurations. As an example, Figure~\ref{fig3}(a) shows another magnetic field map recorded with the scanning-NV magnetometer operated in dual-iso-B imaging mode. Here the three different propagation directions of the antiferromagnetic cycloid can be simultaneously observed in a single ferroelectric domain. Each cycloid variant was identified by considering the orientation of the magnetic field modulation with respect to the in-plane component of the polarization [Fig.~\ref{fig1}(c-e)]. In addition, the period of the modulation was inferred for each magnetic domain by fitting line profiles with a cosine function, leading to $\Lambda_{1}=\SI[multi-part-units=single]{59\pm2}{\nano\meter}$, $\Lambda_{2}=\SI[multi-part-units=single]{82\pm2}{\nano\meter}$, and $\Lambda_{3}=\SI[multi-part-units=single]{94\pm2}{\nano\meter}$ [Fig.~\ref{fig3}(b-d)]. These values are in fair agreement with the expected projections of the cycloid wavevectors on the (001) surface plane. Such a multi-$\mathbf{k}$ domain structure is in stark contrast with the magnetic order observed in \BFO thin films, in which a magnetic anisotropy resulting from epitaxial strain lifts the degeneracy between the three possible propagation directions, leading to the stabilization of a single cycloid variant~\cite{sandoCraftingMagnonicSpintronic2013, haykalAntiferromagneticTexturesBiFeO32020}.

Earlier high-resolution neutron diffraction experiments performed on bulk \BFO crystals have also revealed the coexistence of the three cycloid variants, with a predominance of $\mathbf{k_1}$ domains~\cite{ramazanogluLocalWeakFerromagnetism2011, ramazanogluTemperature2011}. 
However, we note that for single crystals from the same batch as the ones used in this work, a single cycloidal propagation direction along $\mathbf{k_1}$ was detected by neutron diffraction~\cite{lebeugleElectricFieldInducedSpinFlop2008}. Thus, the nucleation and growth of a single spin cycloid among the three possible ones seems to be favored in the core of the crystal, and then split into a more complex multi-$\mathbf{k}$ domain structure at the crystal surface. Indeed, while neutron diffraction probes the full sample volume, scanning-NV magnetometry is only sensitive to stray fields generated by the last 30-\SI{50}{\nano\meter} of the crystal from its surface. To reconcile bulk and surface observations, we can assume that surface symmetry breaking is associated to a depolarizing electrical field, which might perturb the cycloidal ordering through the magnetoelectric coupling. Indeed a not fully screened polarization would produce an extra (negative) electric field component along [001]. Because the magnetoelectric effect favors cycloids with magnetic moments contained in their $(\mathbf{E},\mathbf{k})$ plane and propagation vectors perpendicular to the total electric field, our (001) surface would weaken the $\mathbf{k_1}$ cycloidal state compared to $\mathbf{k_2}$ and $\mathbf{k_3}$ (affected equally). Therefore, a bulk $\mathbf{k_1}$ cycloid would be destabilized at the surface and could thus produce the multi-$\mathbf{k}$ domain structure observed here. Further theoretical inputs are highly desirable to explore this issue.

\begin{figure*}[t]
\includegraphics[width=17.5cm]{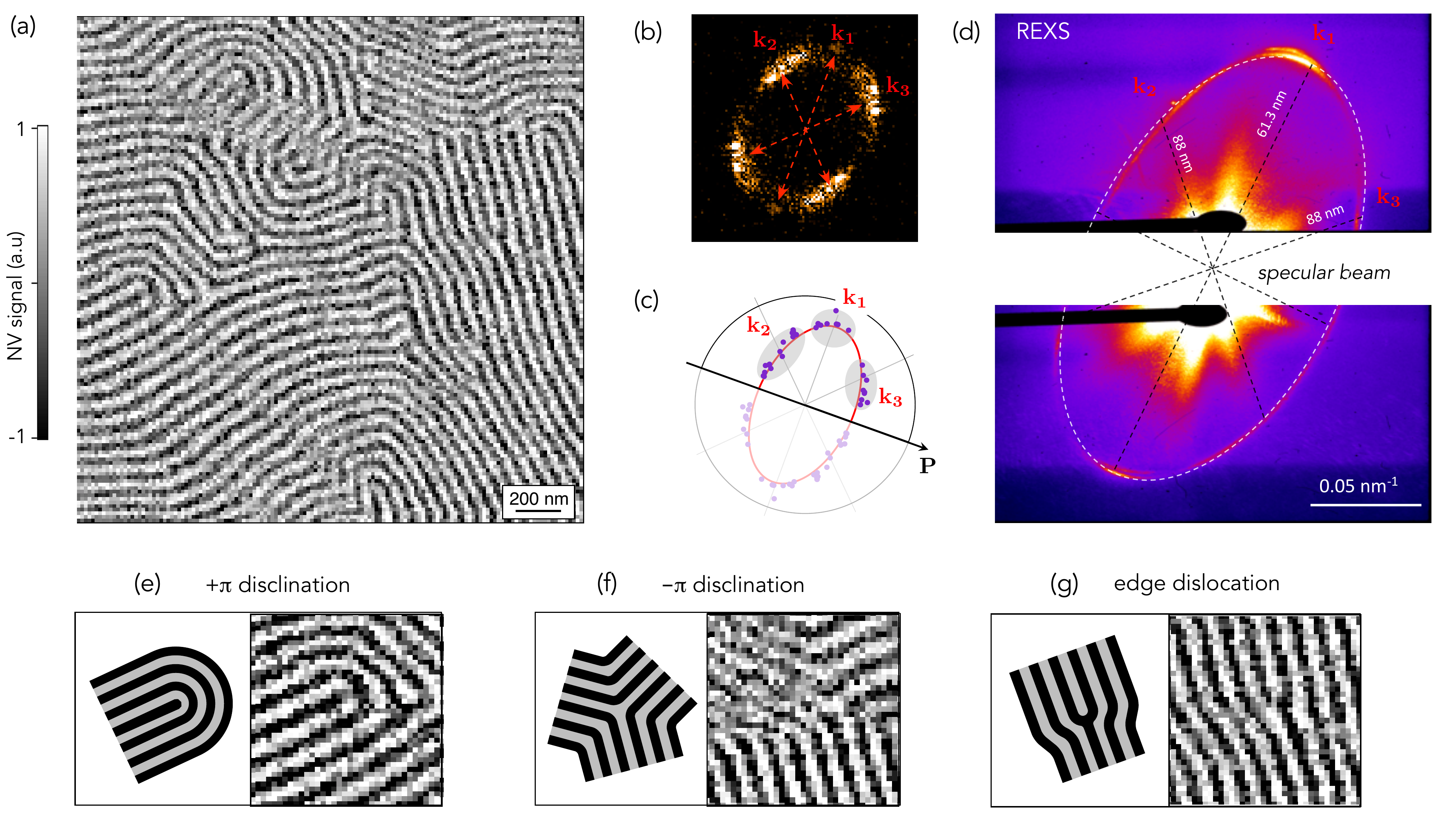}
\caption{(a) NV magnetometry image recorded in dual-iso-B mode showing a continuous rotation of the cycloid propagation direction. (b) Fourier transform of the magnetic image shown in (a). The red dashed arrows indicate the three expected cycloid wavevectors. (c) Polar diagram of the in-plane projection of the wavevector obtained by measuring the period of the magnetic field modulation and its orientation with respect to the polarization in different regions of the sample. The red solid line corresponds to the projection on the surface plane of the circle with radius $2\pi/\lambda_b$ in the (111) plane. The shaded area of the diagram is reconstructed by symmetry. (d) Resonant elastic X-ray scattering (REXS) at the Fe $L$-edge (\SI{708.78}{\electronvolt}). The elliptic diffraction pattern further confirms the continuous rotation of the cycloid propagation directions in the (111) plane. (e-g) Sketches and corresponding NV magnetometry close views of the different topological defects found in bulk \BFO: $\pm\pi$-disclinations and edge dislocations.}
\label{fig4}
\end{figure*}

The magnetic image shown in Figure~\ref{fig3}(a) indicates that the transition between the magnetic domains is realized either via a smooth rotation of the cycloidal wavevector or through the formation of complex whirling structures. This is further illustrated by Figure~\ref{fig4}(a) showing another magnetic image recorded in a different region of the crystal. Here the propagation direction of the cycloid is continuously rotating such that the boundaries of magnetic domains can hardly be identified. The Fourier transform of this magnetic image features an elliptic shape, which illustrates that the continuous rotation of the cycloid wavevector is correlated with a variation of the surface magnetic modulation period [Fig.~\ref{fig4}(b)]. The major axis of the ellipse corresponds to the in-plane propagation direction $\mathbf{k_1}$, while the two diffused satellites centered at $\pm \ang{45}$ from this axis are associated to $\mathbf{k_2}$ and $\mathbf{k_3}$. We note that the spots corresponding to $\mathbf{k_1}$ are less intense because of the dominance of the out-of-plane variants of the cycloid in the magnetic image of this specific area.

To get a larger scale insight on this cycloid propagation rotation, measurements of the wavevector's norm and orientation with respect to the ferroelectric polarization were realized in different regions of the sample. The results are gathered in the polar diagram shown in Figure~\ref{fig4}(c). The plotted ellipse (red solid line) corresponds to the projection on the surface plane of a circle with radius $2\pi/\lambda_b$ in the (111) plane (see Fig.~\ref{fig1}(a) and Supplemental Material~\cite{suppl}). Interestingly, all experimental data points fall along this ellipse, with a dispersion reflecting the large spread of cycloid propagation directions. This rotation of the cycloid propagation direction in the (111) plane, which was also noticed in neutron diffraction experiments~\cite{sosnowskaSpiralMagneticOrdering1982, ramazanogluLocalWeakFerromagnetism2011}, results from the connection between the different rotational magnetic domains. Complementary measurements were performed in reciprocal space using resonant elastic X-ray scattering (REXS) at the Fe $L$-edge~\cite{johnsonXRayImagingMultiferroic2013,waterfieldpriceCoherentMagnetoelasticDomains2016}, allowing a large surface scale to be integrated (typically of the order of \SI{100}{\micro\meter} compared to \SI{1}{\micro\meter} for scanning-NV magnetometry). This technique, which makes use of soft X-rays in reflectivity geometry, mostly probes the projection of the cycloid period on the surface plane, as the typical absorption length around the Fe $L$-edge is only a few tens of \si{\nano\meter} from the surface of the crystal. Therefore, similarly to the scanning-NV measurements, it is only sensitive to the surface magnetic state. As shown in Figure~\ref{fig4}(d), we obtain an elliptical diffraction pattern from the REXS experiment, which nicely corroborates the results obtained in real space with scanning-NV microscopy. Furthermore, the ellipse long axis of \SI{62}{\nano\meter} corresponds to the in-plane $\mathbf{k_1}$ vector while its short axis of \SI{108}{\nano\meter} aligns in between the $\mathbf{k_2}$ and $\mathbf{k_3}$ vectors. This enhanced projected period at the short axis may indicate that the continuous rotation of the propagation vectors could step by the $[11\overline{2}]$ crystallographic axis, identified as the \emph{type II} cycloid. Indeed, recent experiments have shown that such an alternative cycloid, can be stabilized in epitaxial \BFO thin films~\cite{sandoCraftingMagnonicSpintronic2013, agbeleleStrainMagneticField2017, haykalAntiferromagneticTexturesBiFeO32020, Sando2020}. This exotic cycloidal order appears to have a period similar to the type~I cycloid and can propagate along three possible directions: $\mathbf{k'_1} \parallel [\overline{2}11]$, $\mathbf{k'_2}  \parallel [1\overline{2}1]$ and $\mathbf{k'_3}  \parallel [11\overline{2}]$, which are all contained in the (111) plane~\cite{haykalAntiferromagneticTexturesBiFeO32020}. The continuous rotation of the propagation vector between its three main components in single crystals might then mediated by the exotic cycloidal ordering, as supported by theoretical predictions of multiple propagation vectors in \BFO~\cite{xuRevisitingSpinCycloids2018}.

Going back to the magnetic configuration presented in Fig.~\ref{fig4}(a), we now focus on the boundaries between the different rotational domains, where we expect the formation of the topological defects which are typically found in lamellar structures~\cite{seul1995domain}. Fig.~\ref{fig4}(e)-(g) display selected regions of the larger scale images shown in Figs.~\ref{fig3}(a) and \ref{fig4}(a), where singularities can be identified. Although the lack of complete translational invariance in lamellar systems prevents a rigorous topological classification of these defects~\cite{merminTopologicalTheoryDefects1979}, the naive generalization of this approach allows us to describe such singularities, considering that the order parameter is the cycloid propagation direction. The textures presented in Figs~\ref{fig4}(e) and (f) are $\pm \pi$-disclinations, as the cycloid wavevector direction covers only half a circle (rotating in opposite senses for the two defects) when we follow a contour enclosing the singularity. The edge dislocation shown in Fig.~\ref{fig4}(g) corresponds to a combination of the $+\pi$ and $-\pi$-disclinations. Such edge dislocations have a winding number of 0 or $\pm \nicefrac{1}{2}$ depending on the distance between the two disclinations~\cite{schoenherrTopologicalDomainWalls2018a, schonherrTopologicalStructuresMagnetic2018}. Albeit all these defects do not have a well-defined or finite integer winding number, meaning that topology does not prevent their annihilation, they are stabilized by the cycloidal lamellar structure resulting from the interplay between the different magnetic interactions in bulk \BFO.
Our NV magnetometry images of the topological defects are strikingly similar to the MFM images obtained in the chiral ferromagnet FeGe~\cite{schoenherrTopologicalDomainWalls2018a}, even if they do not arrange along straight domain walls separating regions where the magnetic spiral propagates along different directions. As discussed previously from the large spread of the data points on the ellipse in Figs.~\ref{fig4}(c) and (d), it appears that the cycloid wavevector in bulk \BFO is not stricly locked to the expected crystallographic directions at the vicinity of the crystal surface. This weak anisotropy allows both the smooth transitions between the rotational domains and the formation of isolated defects.


To conclude, we have studied the cycloidal antiferromagnetic order at the surface of a bulk \BFO crystal. Combining reciprocal and real-space magnetic imaging techniques, we have shown the coexistence of antiferromagnetic domains in which the cycloid propagates along different wavevectors. In addition, owing to the nanoscale spatial resolution of scanning NV magnetometry, we observed the formation of topological defects typical for lamellar materials at the junctions between these domains. Our work thus demonstrates that these magnetic objects, previously observed in a chiral ferromagnet~\cite{schoenherrTopologicalDomainWalls2018a}, can be transposed in a multiferroic antiferromagnet, offering new opportunities in terms of robustness and electrical control towards their use in spintronic devices. \\

\noindent {\it Acknowledgments} -  We thank support from the French Agence Nationale de la Recherche (ANR) through the project TATOO (ANR-21-CE09-0033-01), the European Union’s Horizon 2020 research and innovation programme under grant agreement No. 964931 (TSAR) and No. 866267 (EXAFONIS), the DARPA TEE program and a public grant overseen by the ANR as part of the ‘Investissement d’Avenir’ programme (LABEX NanoSaclay, ref. ANR-10-LABX-0035, SPiCY). The Sesame Ile de France IMAGeSPIN project (No. EX039175) is also acknowledged.


\begin{thebibliography}{32}%
\makeatletter
\providecommand \@ifxundefined [1]{%
 \@ifx{#1\undefined}
}%
\providecommand \@ifnum [1]{%
 \ifnum #1\expandafter \@firstoftwo
 \else \expandafter \@secondoftwo
 \fi
}%
\providecommand \@ifx [1]{%
 \ifx #1\expandafter \@firstoftwo
 \else \expandafter \@secondoftwo
 \fi
}%
\providecommand \natexlab [1]{#1}%
\providecommand \enquote  [1]{``#1''}%
\providecommand \bibnamefont  [1]{#1}%
\providecommand \bibfnamefont [1]{#1}%
\providecommand \citenamefont [1]{#1}%
\providecommand \href@noop [0]{\@secondoftwo}%
\providecommand \href [0]{\begingroup \@sanitize@url \@href}%
\providecommand \@href[1]{\@@startlink{#1}\@@href}%
\providecommand \@@href[1]{\endgroup#1\@@endlink}%
\providecommand \@sanitize@url [0]{\catcode `\\12\catcode `\$12\catcode
  `\&12\catcode `\#12\catcode `\^12\catcode `\_12\catcode `\%12\relax}%
\providecommand \@@startlink[1]{}%
\providecommand \@@endlink[0]{}%
\providecommand \url  [0]{\begingroup\@sanitize@url \@url }%
\providecommand \@url [1]{\endgroup\@href {#1}{\urlprefix }}%
\providecommand \urlprefix  [0]{URL }%
\providecommand \Eprint [0]{\href }%
\providecommand \doibase [0]{https://doi.org/}%
\providecommand \selectlanguage [0]{\@gobble}%
\providecommand \bibinfo  [0]{\@secondoftwo}%
\providecommand \bibfield  [0]{\@secondoftwo}%
\providecommand \translation [1]{[#1]}%
\providecommand \BibitemOpen [0]{}%
\providecommand \bibitemStop [0]{}%
\providecommand \bibitemNoStop [0]{.\EOS\space}%
\providecommand \EOS [0]{\spacefactor3000\relax}%
\providecommand \BibitemShut  [1]{\csname bibitem#1\endcsname}%
\let\auto@bib@innerbib\@empty
\bibitem [{\citenamefont
  {Zurek}(1996)}]{zurekCosmologicalExperimentsCondensed1996}%
  \BibitemOpen
  \bibfield  {author} {\bibinfo {author} {\bibfnamefont {W.~H.}\ \bibnamefont
  {Zurek}},\ }\href {https://doi.org/10.1016/S0370-1573(96)00009-9} {\bibfield
  {journal} {\bibinfo  {journal} {Physics Reports}\ }\textbf {\bibinfo {volume}
  {276}},\ \bibinfo {pages} {177} (\bibinfo {year} {1996})}\BibitemShut
  {NoStop}%
\bibitem [{\citenamefont {Mermin}(1979)}]{merminTopologicalTheoryDefects1979}%
  \BibitemOpen
  \bibfield  {author} {\bibinfo {author} {\bibfnamefont {N.~D.}\ \bibnamefont
  {Mermin}},\ }\href {https://doi.org/10.1103/RevModPhys.51.591} {\bibfield
  {journal} {\bibinfo  {journal} {Reviews of Modern Physics}\ }\textbf
  {\bibinfo {volume} {51}},\ \bibinfo {pages} {591} (\bibinfo {year}
  {1979})}\BibitemShut {NoStop}%
\bibitem [{\citenamefont {de~Gennes}\ and\ \citenamefont
  {Prost}(1993)}]{deGennesLiquidCrystals}%
  \BibitemOpen
  \bibfield  {author} {\bibinfo {author} {\bibfnamefont {P.-G.}\ \bibnamefont
  {de~Gennes}}\ and\ \bibinfo {author} {\bibfnamefont {J.}~\bibnamefont
  {Prost}},\ }\href@noop {} {\emph {\bibinfo {title} {The Physics of Liquid
  Crystals}}}\ (\bibinfo  {publisher} {Clarendon Press, Oxford},\ \bibinfo
  {year} {1993})\BibitemShut {NoStop}%
\bibitem [{\citenamefont {Seul}\ and\ \citenamefont
  {Andelman}(1995)}]{seul1995domain}%
  \BibitemOpen
  \bibfield  {author} {\bibinfo {author} {\bibfnamefont {M.}~\bibnamefont
  {Seul}}\ and\ \bibinfo {author} {\bibfnamefont {D.}~\bibnamefont
  {Andelman}},\ }\href {https://doi.org/10.1126/science.267.5197.476}
  {\bibfield  {journal} {\bibinfo  {journal} {Science}\ }\textbf {\bibinfo
  {volume} {267}},\ \bibinfo {pages} {476} (\bibinfo {year}
  {1995})}\BibitemShut {NoStop}%
\bibitem [{\citenamefont {Bouligand}(1983)}]{Bouligand}%
  \BibitemOpen
  \bibfield  {author} {\bibinfo {author} {\bibfnamefont {Y.}~\bibnamefont
  {Bouligand}},\ }\href@noop {} {\emph {\bibinfo {title} {Dislocations in
  solids, Chap. 23}}}\ (\bibinfo  {publisher} {North-Holland Publishing
  Company},\ \bibinfo {year} {1983})\BibitemShut {NoStop}%
\bibitem [{\citenamefont {Seul}\ and\ \citenamefont
  {Wolfe}(1992)}]{seul1992evolution}%
  \BibitemOpen
  \bibfield  {author} {\bibinfo {author} {\bibfnamefont {M.}~\bibnamefont
  {Seul}}\ and\ \bibinfo {author} {\bibfnamefont {R.}~\bibnamefont {Wolfe}},\
  }\href {https://doi.org/10.1103/PhysRevA.46.7519} {\bibfield  {journal}
  {\bibinfo  {journal} {Physical Review A}\ }\textbf {\bibinfo {volume} {46}},\
  \bibinfo {pages} {7519} (\bibinfo {year} {1992})}\BibitemShut {NoStop}%
\bibitem [{\citenamefont {Murphy}\ \emph {et~al.}(2015)\citenamefont {Murphy},
  \citenamefont {Harris},\ and\ \citenamefont {Buriak}}]{murphy2015automated}%
  \BibitemOpen
  \bibfield  {author} {\bibinfo {author} {\bibfnamefont {J.~N.}\ \bibnamefont
  {Murphy}}, \bibinfo {author} {\bibfnamefont {K.~D.}\ \bibnamefont {Harris}},\
  and\ \bibinfo {author} {\bibfnamefont {J.~M.}\ \bibnamefont {Buriak}},\
  }\href {https://doi.org/10.1371/journal.pone.0133088} {\bibfield  {journal}
  {\bibinfo  {journal} {PLoS One}\ }\textbf {\bibinfo {volume} {10}},\ \bibinfo
  {pages} {e0133088} (\bibinfo {year} {2015})}\BibitemShut {NoStop}%
\bibitem [{\citenamefont {Schoenherr}\ \emph {et~al.}(2018)\citenamefont
  {Schoenherr}, \citenamefont {Müller}, \citenamefont {Köhler}, \citenamefont
  {Rosch}, \citenamefont {Kanazawa}, \citenamefont {Tokura}, \citenamefont
  {Garst},\ and\ \citenamefont
  {Meier}}]{schoenherrTopologicalDomainWalls2018a}%
  \BibitemOpen
  \bibfield  {author} {\bibinfo {author} {\bibfnamefont {P.}~\bibnamefont
  {Schoenherr}}, \bibinfo {author} {\bibfnamefont {J.}~\bibnamefont {Müller}},
  \bibinfo {author} {\bibfnamefont {L.}~\bibnamefont {Köhler}}, \bibinfo
  {author} {\bibfnamefont {A.}~\bibnamefont {Rosch}}, \bibinfo {author}
  {\bibfnamefont {N.}~\bibnamefont {Kanazawa}}, \bibinfo {author}
  {\bibfnamefont {Y.}~\bibnamefont {Tokura}}, \bibinfo {author} {\bibfnamefont
  {M.}~\bibnamefont {Garst}},\ and\ \bibinfo {author} {\bibfnamefont
  {D.}~\bibnamefont {Meier}},\ }\href
  {https://doi.org/10.1038/s41567-018-0056-5} {\bibfield  {journal} {\bibinfo
  {journal} {Nature Physics}\ }\textbf {\bibinfo {volume} {14}},\ \bibinfo
  {pages} {465} (\bibinfo {year} {2018})}\BibitemShut {NoStop}%
\bibitem [{\citenamefont {Yu}\ \emph {et~al.}(2011)\citenamefont {Yu},
  \citenamefont {Kanazawa}, \citenamefont {Onose}, \citenamefont {Kimoto},
  \citenamefont {Zhang}, \citenamefont {Ishiwata}, \citenamefont {Matsui},\
  and\ \citenamefont {Tokura}}]{yuRoomtemperatureFormationSkyrmion2011}%
  \BibitemOpen
  \bibfield  {author} {\bibinfo {author} {\bibfnamefont {X.~Z.}\ \bibnamefont
  {Yu}}, \bibinfo {author} {\bibfnamefont {N.}~\bibnamefont {Kanazawa}},
  \bibinfo {author} {\bibfnamefont {Y.}~\bibnamefont {Onose}}, \bibinfo
  {author} {\bibfnamefont {K.}~\bibnamefont {Kimoto}}, \bibinfo {author}
  {\bibfnamefont {W.~Z.}\ \bibnamefont {Zhang}}, \bibinfo {author}
  {\bibfnamefont {S.}~\bibnamefont {Ishiwata}}, \bibinfo {author}
  {\bibfnamefont {Y.}~\bibnamefont {Matsui}},\ and\ \bibinfo {author}
  {\bibfnamefont {Y.}~\bibnamefont {Tokura}},\ }\href
  {https://doi.org/10.1038/nmat2916} {\bibfield  {journal} {\bibinfo  {journal}
  {Nature Materials}\ }\textbf {\bibinfo {volume} {10}},\ \bibinfo {pages}
  {106} (\bibinfo {year} {2011})}\BibitemShut {NoStop}%
\bibitem [{\citenamefont {Sosnowska}\ \emph {et~al.}(1982)\citenamefont
  {Sosnowska}, \citenamefont {Neumaier},\ and\ \citenamefont
  {Steichele}}]{sosnowskaSpiralMagneticOrdering1982}%
  \BibitemOpen
  \bibfield  {author} {\bibinfo {author} {\bibfnamefont {I.}~\bibnamefont
  {Sosnowska}}, \bibinfo {author} {\bibfnamefont {T.~P.}\ \bibnamefont
  {Neumaier}},\ and\ \bibinfo {author} {\bibfnamefont {E.}~\bibnamefont
  {Steichele}},\ }\href {https://doi.org/10.1088/0022-3719/15/23/020}
  {\bibfield  {journal} {\bibinfo  {journal} {J. Phys. C: Solid State Phys.}\
  }\textbf {\bibinfo {volume} {15}},\ \bibinfo {pages} {4835} (\bibinfo {year}
  {1982})}\BibitemShut {NoStop}%
\bibitem [{\citenamefont {Moreau}\ \emph {et~al.}(1971)\citenamefont {Moreau},
  \citenamefont {Michel}, \citenamefont {Gerson},\ and\ \citenamefont
  {James}}]{moreauFerroelectricBiFeO3Xray1971}%
  \BibitemOpen
  \bibfield  {author} {\bibinfo {author} {\bibfnamefont {J.~M.}\ \bibnamefont
  {Moreau}}, \bibinfo {author} {\bibfnamefont {C.}~\bibnamefont {Michel}},
  \bibinfo {author} {\bibfnamefont {R.}~\bibnamefont {Gerson}},\ and\ \bibinfo
  {author} {\bibfnamefont {W.~J.}\ \bibnamefont {James}},\ }\href
  {https://doi.org/10.1016/S0022-3697(71)80189-0} {\bibfield  {journal}
  {\bibinfo  {journal} {Journal of Physics and Chemistry of Solids}\ }\textbf
  {\bibinfo {volume} {32}},\ \bibinfo {pages} {1315} (\bibinfo {year}
  {1971})}\BibitemShut {NoStop}%
\bibitem [{\citenamefont {Kubel}\ and\ \citenamefont
  {Schmid}(1990)}]{kubelStructureFerroelectricFerroelastic1990}%
  \BibitemOpen
  \bibfield  {author} {\bibinfo {author} {\bibfnamefont {F.}~\bibnamefont
  {Kubel}}\ and\ \bibinfo {author} {\bibfnamefont {H.}~\bibnamefont {Schmid}},\
  }\href {https://doi.org/10.1107/S0108768190006887} {\bibfield  {journal}
  {\bibinfo  {journal} {Acta Crystallographica Section B}\ }\textbf {\bibinfo
  {volume} {46}},\ \bibinfo {pages} {698} (\bibinfo {year} {1990})}\BibitemShut
  {NoStop}%
\bibitem [{\citenamefont {Wang}\ \emph {et~al.}(2003)\citenamefont {Wang},
  \citenamefont {Neaton}, \citenamefont {Zheng}, \citenamefont {Nagarajan},
  \citenamefont {Ogale}, \citenamefont {Liu}, \citenamefont {Viehland},
  \citenamefont {Vaithyanathan}, \citenamefont {Schlom}, \citenamefont
  {Waghmare}, \citenamefont {Spaldin}, \citenamefont {Rabe}, \citenamefont
  {Wuttig},\ and\ \citenamefont
  {Ramesh}}]{wangEpitaxialBiFeO3Multiferroic2003}%
  \BibitemOpen
  \bibfield  {author} {\bibinfo {author} {\bibfnamefont {J.}~\bibnamefont
  {Wang}}, \bibinfo {author} {\bibfnamefont {J.~B.}\ \bibnamefont {Neaton}},
  \bibinfo {author} {\bibfnamefont {H.}~\bibnamefont {Zheng}}, \bibinfo
  {author} {\bibfnamefont {V.}~\bibnamefont {Nagarajan}}, \bibinfo {author}
  {\bibfnamefont {S.~B.}\ \bibnamefont {Ogale}}, \bibinfo {author}
  {\bibfnamefont {B.}~\bibnamefont {Liu}}, \bibinfo {author} {\bibfnamefont
  {D.}~\bibnamefont {Viehland}}, \bibinfo {author} {\bibfnamefont
  {V.}~\bibnamefont {Vaithyanathan}}, \bibinfo {author} {\bibfnamefont {D.~G.}\
  \bibnamefont {Schlom}}, \bibinfo {author} {\bibfnamefont {U.~V.}\
  \bibnamefont {Waghmare}}, \bibinfo {author} {\bibfnamefont {N.~A.}\
  \bibnamefont {Spaldin}}, \bibinfo {author} {\bibfnamefont {K.~M.}\
  \bibnamefont {Rabe}}, \bibinfo {author} {\bibfnamefont {M.}~\bibnamefont
  {Wuttig}},\ and\ \bibinfo {author} {\bibfnamefont {R.}~\bibnamefont
  {Ramesh}},\ }\href {https://doi.org/10.1126/science.1080615} {\bibfield
  {journal} {\bibinfo  {journal} {Science}\ }\textbf {\bibinfo {volume}
  {299}},\ \bibinfo {pages} {1719} (\bibinfo {year} {2003})}\BibitemShut
  {NoStop}%
\bibitem [{\citenamefont {Lebeugle}\ \emph {et~al.}(2007)\citenamefont
  {Lebeugle}, \citenamefont {Colson}, \citenamefont {Forget}, \citenamefont
  {Viret}, \citenamefont {Bonville}, \citenamefont {Marucco},\ and\
  \citenamefont {Fusil}}]{lebeugleRoomtemperatureCoexistenceLarge2007}%
  \BibitemOpen
  \bibfield  {author} {\bibinfo {author} {\bibfnamefont {D.}~\bibnamefont
  {Lebeugle}}, \bibinfo {author} {\bibfnamefont {D.}~\bibnamefont {Colson}},
  \bibinfo {author} {\bibfnamefont {A.}~\bibnamefont {Forget}}, \bibinfo
  {author} {\bibfnamefont {M.}~\bibnamefont {Viret}}, \bibinfo {author}
  {\bibfnamefont {P.}~\bibnamefont {Bonville}}, \bibinfo {author}
  {\bibfnamefont {J.~F.}\ \bibnamefont {Marucco}},\ and\ \bibinfo {author}
  {\bibfnamefont {S.}~\bibnamefont {Fusil}},\ }\href
  {https://doi.org/10.1103/PhysRevB.76.024116} {\bibfield  {journal} {\bibinfo
  {journal} {Phys. Rev. B}\ }\textbf {\bibinfo {volume} {76}},\ \bibinfo
  {pages} {024116} (\bibinfo {year} {2007})}\BibitemShut {NoStop}%
\bibitem [{\citenamefont {Katsura}\ \emph {et~al.}(2005)\citenamefont
  {Katsura}, \citenamefont {Nagaosa},\ and\ \citenamefont
  {Balatsky}}]{katsuraSpinCurrentMagnetoelectric2005}%
  \BibitemOpen
  \bibfield  {author} {\bibinfo {author} {\bibfnamefont {H.}~\bibnamefont
  {Katsura}}, \bibinfo {author} {\bibfnamefont {N.}~\bibnamefont {Nagaosa}},\
  and\ \bibinfo {author} {\bibfnamefont {A.~V.}\ \bibnamefont {Balatsky}},\
  }\href {https://doi.org/10.1103/PhysRevLett.95.057205} {\bibfield  {journal}
  {\bibinfo  {journal} {Phys. Rev. Lett.}\ }\textbf {\bibinfo {volume} {95}},\
  \bibinfo {pages} {057205} (\bibinfo {year} {2005})}\BibitemShut {NoStop}%
\bibitem [{\citenamefont {Lebeugle}\ \emph {et~al.}(2008)\citenamefont
  {Lebeugle}, \citenamefont {Colson}, \citenamefont {Forget}, \citenamefont
  {Viret}, \citenamefont {Bataille},\ and\ \citenamefont
  {Gukasov}}]{lebeugleElectricFieldInducedSpinFlop2008}%
  \BibitemOpen
  \bibfield  {author} {\bibinfo {author} {\bibfnamefont {D.}~\bibnamefont
  {Lebeugle}}, \bibinfo {author} {\bibfnamefont {D.}~\bibnamefont {Colson}},
  \bibinfo {author} {\bibfnamefont {A.}~\bibnamefont {Forget}}, \bibinfo
  {author} {\bibfnamefont {M.}~\bibnamefont {Viret}}, \bibinfo {author}
  {\bibfnamefont {A.~M.}\ \bibnamefont {Bataille}},\ and\ \bibinfo {author}
  {\bibfnamefont {A.}~\bibnamefont {Gukasov}},\ }\href
  {https://doi.org/10.1103/PhysRevLett.100.227602} {\bibfield  {journal}
  {\bibinfo  {journal} {Phys. Rev. Lett.}\ }\textbf {\bibinfo {volume} {100}},\
  \bibinfo {pages} {227602} (\bibinfo {year} {2008})}\BibitemShut {NoStop}%
\bibitem [{\citenamefont {Ramazanoglu}\ \emph
  {et~al.}(2011{\natexlab{a}})\citenamefont {Ramazanoglu}, \citenamefont
  {Laver}, \citenamefont {Ratcliff}, \citenamefont {Watson}, \citenamefont
  {Chen}, \citenamefont {Jackson}, \citenamefont {Kothapalli}, \citenamefont
  {Lee}, \citenamefont {Cheong},\ and\ \citenamefont
  {Kiryukhin}}]{ramazanogluLocalWeakFerromagnetism2011}%
  \BibitemOpen
  \bibfield  {author} {\bibinfo {author} {\bibfnamefont {M.}~\bibnamefont
  {Ramazanoglu}}, \bibinfo {author} {\bibfnamefont {M.}~\bibnamefont {Laver}},
  \bibinfo {author} {\bibfnamefont {W.}~\bibnamefont {Ratcliff}}, \bibinfo
  {author} {\bibfnamefont {S.~M.}\ \bibnamefont {Watson}}, \bibinfo {author}
  {\bibfnamefont {W.~C.}\ \bibnamefont {Chen}}, \bibinfo {author}
  {\bibfnamefont {A.}~\bibnamefont {Jackson}}, \bibinfo {author} {\bibfnamefont
  {K.}~\bibnamefont {Kothapalli}}, \bibinfo {author} {\bibfnamefont
  {S.}~\bibnamefont {Lee}}, \bibinfo {author} {\bibfnamefont {S.-W.}\
  \bibnamefont {Cheong}},\ and\ \bibinfo {author} {\bibfnamefont
  {V.}~\bibnamefont {Kiryukhin}},\ }\href
  {https://doi.org/10.1103/PhysRevLett.107.207206} {\bibfield  {journal}
  {\bibinfo  {journal} {Phys. Rev. Lett.}\ }\textbf {\bibinfo {volume} {107}},\
  \bibinfo {pages} {207206} (\bibinfo {year} {2011}{\natexlab{a}})}\BibitemShut
  {NoStop}%
\bibitem [{\citenamefont {Sando}\ \emph {et~al.}(2013)\citenamefont {Sando},
  \citenamefont {Agbelele}, \citenamefont {Rahmedov}, \citenamefont {Liu},
  \citenamefont {Rovillain}, \citenamefont {Toulouse}, \citenamefont {Infante},
  \citenamefont {Pyatakov}, \citenamefont {Fusil}, \citenamefont {Jacquet},
  \citenamefont {Carr{\'e}t{\'e}ro}, \citenamefont {Deranlot}, \citenamefont
  {Lisenkov}, \citenamefont {Wang}, \citenamefont {Le~Breton}, \citenamefont
  {Cazayous}, \citenamefont {Sacuto}, \citenamefont {Juraszek}, \citenamefont
  {Zvezdin}, \citenamefont {Bellaiche}, \citenamefont {Dkhil}, \citenamefont
  {Barth{\'e}l{\'e}my},\ and\ \citenamefont
  {Bibes}}]{sandoCraftingMagnonicSpintronic2013}%
  \BibitemOpen
  \bibfield  {author} {\bibinfo {author} {\bibfnamefont {D.}~\bibnamefont
  {Sando}}, \bibinfo {author} {\bibfnamefont {A.}~\bibnamefont {Agbelele}},
  \bibinfo {author} {\bibfnamefont {D.}~\bibnamefont {Rahmedov}}, \bibinfo
  {author} {\bibfnamefont {J.}~\bibnamefont {Liu}}, \bibinfo {author}
  {\bibfnamefont {P.}~\bibnamefont {Rovillain}}, \bibinfo {author}
  {\bibfnamefont {C.}~\bibnamefont {Toulouse}}, \bibinfo {author}
  {\bibfnamefont {I.~C.}\ \bibnamefont {Infante}}, \bibinfo {author}
  {\bibfnamefont {A.~P.}\ \bibnamefont {Pyatakov}}, \bibinfo {author}
  {\bibfnamefont {S.}~\bibnamefont {Fusil}}, \bibinfo {author} {\bibfnamefont
  {E.}~\bibnamefont {Jacquet}}, \bibinfo {author} {\bibfnamefont
  {C.}~\bibnamefont {Carr{\'e}t{\'e}ro}}, \bibinfo {author} {\bibfnamefont
  {C.}~\bibnamefont {Deranlot}}, \bibinfo {author} {\bibfnamefont
  {S.}~\bibnamefont {Lisenkov}}, \bibinfo {author} {\bibfnamefont
  {D.}~\bibnamefont {Wang}}, \bibinfo {author} {\bibfnamefont {J.-M.}\
  \bibnamefont {Le~Breton}}, \bibinfo {author} {\bibfnamefont {M.}~\bibnamefont
  {Cazayous}}, \bibinfo {author} {\bibfnamefont {A.}~\bibnamefont {Sacuto}},
  \bibinfo {author} {\bibfnamefont {J.}~\bibnamefont {Juraszek}}, \bibinfo
  {author} {\bibfnamefont {A.~K.}\ \bibnamefont {Zvezdin}}, \bibinfo {author}
  {\bibfnamefont {L.}~\bibnamefont {Bellaiche}}, \bibinfo {author}
  {\bibfnamefont {B.}~\bibnamefont {Dkhil}}, \bibinfo {author} {\bibfnamefont
  {A.}~\bibnamefont {Barth{\'e}l{\'e}my}},\ and\ \bibinfo {author}
  {\bibfnamefont {M.}~\bibnamefont {Bibes}},\ }\href
  {https://doi.org/10.1038/nmat3629} {\bibfield  {journal} {\bibinfo  {journal}
  {Nature Materials}\ }\textbf {\bibinfo {volume} {12}},\ \bibinfo {pages}
  {641} (\bibinfo {year} {2013})}\BibitemShut {NoStop}%
\bibitem [{\citenamefont {Agbelele}\ \emph {et~al.}(2017)\citenamefont
  {Agbelele}, \citenamefont {Sando}, \citenamefont {Toulouse}, \citenamefont
  {Paillard}, \citenamefont {Johnson}, \citenamefont {R{\"u}ffer},
  \citenamefont {Popkov}, \citenamefont {Carr{\'e}t{\'e}ro}, \citenamefont
  {Rovillain}, \citenamefont {Breton}, \citenamefont {Dkhil}, \citenamefont
  {Cazayous}, \citenamefont {Gallais}, \citenamefont {M{\'e}asson},
  \citenamefont {Sacuto}, \citenamefont {Manuel}, \citenamefont {Zvezdin},
  \citenamefont {Barth{\'e}l{\'e}my}, \citenamefont {Juraszek},\ and\
  \citenamefont {Bibes}}]{agbeleleStrainMagneticField2017}%
  \BibitemOpen
  \bibfield  {author} {\bibinfo {author} {\bibfnamefont {A.}~\bibnamefont
  {Agbelele}}, \bibinfo {author} {\bibfnamefont {D.}~\bibnamefont {Sando}},
  \bibinfo {author} {\bibfnamefont {C.}~\bibnamefont {Toulouse}}, \bibinfo
  {author} {\bibfnamefont {C.}~\bibnamefont {Paillard}}, \bibinfo {author}
  {\bibfnamefont {R.~D.}\ \bibnamefont {Johnson}}, \bibinfo {author}
  {\bibfnamefont {R.}~\bibnamefont {R{\"u}ffer}}, \bibinfo {author}
  {\bibfnamefont {A.~F.}\ \bibnamefont {Popkov}}, \bibinfo {author}
  {\bibfnamefont {C.}~\bibnamefont {Carr{\'e}t{\'e}ro}}, \bibinfo {author}
  {\bibfnamefont {P.}~\bibnamefont {Rovillain}}, \bibinfo {author}
  {\bibfnamefont {J.-M.~L.}\ \bibnamefont {Breton}}, \bibinfo {author}
  {\bibfnamefont {B.}~\bibnamefont {Dkhil}}, \bibinfo {author} {\bibfnamefont
  {M.}~\bibnamefont {Cazayous}}, \bibinfo {author} {\bibfnamefont
  {Y.}~\bibnamefont {Gallais}}, \bibinfo {author} {\bibfnamefont {M.-A.}\
  \bibnamefont {M{\'e}asson}}, \bibinfo {author} {\bibfnamefont
  {A.}~\bibnamefont {Sacuto}}, \bibinfo {author} {\bibfnamefont
  {P.}~\bibnamefont {Manuel}}, \bibinfo {author} {\bibfnamefont {A.~K.}\
  \bibnamefont {Zvezdin}}, \bibinfo {author} {\bibfnamefont {A.}~\bibnamefont
  {Barth{\'e}l{\'e}my}}, \bibinfo {author} {\bibfnamefont {J.}~\bibnamefont
  {Juraszek}},\ and\ \bibinfo {author} {\bibfnamefont {M.}~\bibnamefont
  {Bibes}},\ }\href {https://doi.org/10.1002/adma.201602327} {\bibfield
  {journal} {\bibinfo  {journal} {Advanced Materials}\ }\textbf {\bibinfo
  {volume} {29}},\ \bibinfo {pages} {1602327} (\bibinfo {year}
  {2017})}\BibitemShut {NoStop}%
\bibitem [{\citenamefont {Haykal}\ \emph {et~al.}(2020)\citenamefont {Haykal},
  \citenamefont {Fischer}, \citenamefont {Akhtar}, \citenamefont {Chauleau},
  \citenamefont {Sando}, \citenamefont {Finco}, \citenamefont {Godel},
  \citenamefont {Birkh{\"o}lzer}, \citenamefont {Carr{\'e}t{\'e}ro},
  \citenamefont {Jaouen}, \citenamefont {Bibes}, \citenamefont {Viret},
  \citenamefont {Fusil}, \citenamefont {Jacques},\ and\ \citenamefont
  {Garcia}}]{haykalAntiferromagneticTexturesBiFeO32020}%
  \BibitemOpen
  \bibfield  {author} {\bibinfo {author} {\bibfnamefont {A.}~\bibnamefont
  {Haykal}}, \bibinfo {author} {\bibfnamefont {J.}~\bibnamefont {Fischer}},
  \bibinfo {author} {\bibfnamefont {W.}~\bibnamefont {Akhtar}}, \bibinfo
  {author} {\bibfnamefont {J.-Y.}\ \bibnamefont {Chauleau}}, \bibinfo {author}
  {\bibfnamefont {D.}~\bibnamefont {Sando}}, \bibinfo {author} {\bibfnamefont
  {A.}~\bibnamefont {Finco}}, \bibinfo {author} {\bibfnamefont
  {F.}~\bibnamefont {Godel}}, \bibinfo {author} {\bibfnamefont {Y.~A.}\
  \bibnamefont {Birkh{\"o}lzer}}, \bibinfo {author} {\bibfnamefont
  {C.}~\bibnamefont {Carr{\'e}t{\'e}ro}}, \bibinfo {author} {\bibfnamefont
  {N.}~\bibnamefont {Jaouen}}, \bibinfo {author} {\bibfnamefont
  {M.}~\bibnamefont {Bibes}}, \bibinfo {author} {\bibfnamefont
  {M.}~\bibnamefont {Viret}}, \bibinfo {author} {\bibfnamefont
  {S.}~\bibnamefont {Fusil}}, \bibinfo {author} {\bibfnamefont
  {V.}~\bibnamefont {Jacques}},\ and\ \bibinfo {author} {\bibfnamefont
  {V.}~\bibnamefont {Garcia}},\ }\href
  {https://doi.org/10.1038/s41467-020-15501-8} {\bibfield  {journal} {\bibinfo
  {journal} {Nat Commun}\ }\textbf {\bibinfo {volume} {11}},\ \bibinfo {pages}
  {1} (\bibinfo {year} {2020})}\BibitemShut {NoStop}%
\bibitem [{\citenamefont {Burns}\ \emph {et~al.}(2020)\citenamefont {Burns},
  \citenamefont {Paull}, \citenamefont {Juraszek}, \citenamefont {Nagarajan},\
  and\ \citenamefont {Sando}}]{Sando2020}%
  \BibitemOpen
  \bibfield  {author} {\bibinfo {author} {\bibfnamefont {S.~R.}\ \bibnamefont
  {Burns}}, \bibinfo {author} {\bibfnamefont {O.}~\bibnamefont {Paull}},
  \bibinfo {author} {\bibfnamefont {J.}~\bibnamefont {Juraszek}}, \bibinfo
  {author} {\bibfnamefont {V.}~\bibnamefont {Nagarajan}},\ and\ \bibinfo
  {author} {\bibfnamefont {D.}~\bibnamefont {Sando}},\ }\href
  {https://doi.org/https://doi.org/10.1002/adma.202003711} {\bibfield
  {journal} {\bibinfo  {journal} {Advanced Materials}\ }\textbf {\bibinfo
  {volume} {32}},\ \bibinfo {pages} {2003711} (\bibinfo {year}
  {2020})}\BibitemShut {NoStop}%
\bibitem [{\citenamefont {Gross}\ \emph {et~al.}(2017)\citenamefont {Gross},
  \citenamefont {Akhtar}, \citenamefont {Garcia}, \citenamefont {Mart{\'i}nez},
  \citenamefont {Chouaieb}, \citenamefont {Garcia}, \citenamefont
  {Carr{\'e}t{\'e}ro}, \citenamefont {Barth{\'e}l{\'e}my}, \citenamefont
  {Appel}, \citenamefont {Maletinsky}, \citenamefont {Kim}, \citenamefont
  {Chauleau}, \citenamefont {Jaouen}, \citenamefont {Viret}, \citenamefont
  {Bibes}, \citenamefont {Fusil},\ and\ \citenamefont
  {Jacques}}]{grossRealspaceImagingNoncollinear2017}%
  \BibitemOpen
  \bibfield  {author} {\bibinfo {author} {\bibfnamefont {I.}~\bibnamefont
  {Gross}}, \bibinfo {author} {\bibfnamefont {W.}~\bibnamefont {Akhtar}},
  \bibinfo {author} {\bibfnamefont {V.}~\bibnamefont {Garcia}}, \bibinfo
  {author} {\bibfnamefont {L.~J.}\ \bibnamefont {Mart{\'i}nez}}, \bibinfo
  {author} {\bibfnamefont {S.}~\bibnamefont {Chouaieb}}, \bibinfo {author}
  {\bibfnamefont {K.}~\bibnamefont {Garcia}}, \bibinfo {author} {\bibfnamefont
  {C.}~\bibnamefont {Carr{\'e}t{\'e}ro}}, \bibinfo {author} {\bibfnamefont
  {A.}~\bibnamefont {Barth{\'e}l{\'e}my}}, \bibinfo {author} {\bibfnamefont
  {P.}~\bibnamefont {Appel}}, \bibinfo {author} {\bibfnamefont
  {P.}~\bibnamefont {Maletinsky}}, \bibinfo {author} {\bibfnamefont {J.-V.}\
  \bibnamefont {Kim}}, \bibinfo {author} {\bibfnamefont {J.~Y.}\ \bibnamefont
  {Chauleau}}, \bibinfo {author} {\bibfnamefont {N.}~\bibnamefont {Jaouen}},
  \bibinfo {author} {\bibfnamefont {M.}~\bibnamefont {Viret}}, \bibinfo
  {author} {\bibfnamefont {M.}~\bibnamefont {Bibes}}, \bibinfo {author}
  {\bibfnamefont {S.}~\bibnamefont {Fusil}},\ and\ \bibinfo {author}
  {\bibfnamefont {V.}~\bibnamefont {Jacques}},\ }\href
  {https://doi.org/10.1038/nature23656} {\bibfield  {journal} {\bibinfo
  {journal} {Nature}\ }\textbf {\bibinfo {volume} {549}},\ \bibinfo {pages}
  {252} (\bibinfo {year} {2017})}\BibitemShut {NoStop}%
\bibitem [{\citenamefont {Chauleau}\ \emph {et~al.}(2020)\citenamefont
  {Chauleau}, \citenamefont {Chirac}, \citenamefont {Fusil}, \citenamefont
  {Garcia}, \citenamefont {Akhtar}, \citenamefont {Tranchida}, \citenamefont
  {Thibaudeau}, \citenamefont {Gross}, \citenamefont {Blouzon}, \citenamefont
  {Finco}, \citenamefont {Bibes}, \citenamefont {Dkhil}, \citenamefont
  {Khalyavin}, \citenamefont {Manuel}, \citenamefont {Jacques}, \citenamefont
  {Jaouen},\ and\ \citenamefont
  {Viret}}]{chauleauElectricAntiferromagneticChiral2020}%
  \BibitemOpen
  \bibfield  {author} {\bibinfo {author} {\bibfnamefont {J.-Y.}\ \bibnamefont
  {Chauleau}}, \bibinfo {author} {\bibfnamefont {T.}~\bibnamefont {Chirac}},
  \bibinfo {author} {\bibfnamefont {S.}~\bibnamefont {Fusil}}, \bibinfo
  {author} {\bibfnamefont {V.}~\bibnamefont {Garcia}}, \bibinfo {author}
  {\bibfnamefont {W.}~\bibnamefont {Akhtar}}, \bibinfo {author} {\bibfnamefont
  {J.}~\bibnamefont {Tranchida}}, \bibinfo {author} {\bibfnamefont
  {P.}~\bibnamefont {Thibaudeau}}, \bibinfo {author} {\bibfnamefont
  {I.}~\bibnamefont {Gross}}, \bibinfo {author} {\bibfnamefont
  {C.}~\bibnamefont {Blouzon}}, \bibinfo {author} {\bibfnamefont
  {A.}~\bibnamefont {Finco}}, \bibinfo {author} {\bibfnamefont
  {M.}~\bibnamefont {Bibes}}, \bibinfo {author} {\bibfnamefont
  {B.}~\bibnamefont {Dkhil}}, \bibinfo {author} {\bibfnamefont {D.~D.}\
  \bibnamefont {Khalyavin}}, \bibinfo {author} {\bibfnamefont {P.}~\bibnamefont
  {Manuel}}, \bibinfo {author} {\bibfnamefont {V.}~\bibnamefont {Jacques}},
  \bibinfo {author} {\bibfnamefont {N.}~\bibnamefont {Jaouen}},\ and\ \bibinfo
  {author} {\bibfnamefont {M.}~\bibnamefont {Viret}},\ }\href
  {https://doi.org/10.1038/s41563-019-0516-z} {\bibfield  {journal} {\bibinfo
  {journal} {Nature Materials}\ }\textbf {\bibinfo {volume} {19}},\ \bibinfo
  {pages} {386} (\bibinfo {year} {2020})}\BibitemShut {NoStop}%
\bibitem [{\citenamefont {Gross}(2017)}]{grossExploringNoncollinearSpin2017}%
  \BibitemOpen
  \bibfield  {author} {\bibinfo {author} {\bibfnamefont {I.}~\bibnamefont
  {Gross}},\ }\emph {\bibinfo {title} {Exploring Non-Collinear Spin Structures
  in Thin Magnetic Films with {{Nitrogen}}-{{Vacancy Scanning}}
  Magnetometry}},\ \href
  {https://tel.archives-ouvertes.fr/tel-01792316/document} {Ph.D. thesis},\
  \bibinfo  {school} {Universit\'e Paris-Saclay} (\bibinfo {year}
  {2017})\BibitemShut {NoStop}%
\bibitem [{sup()}]{suppl}%
  \BibitemOpen
  \href@noop {} {\bibinfo  {journal} {Supplemental Material giving details
  about the analytical calculations of the magnetic stray field distribution
  and about the probe-to-sample distance calculation}\ }\BibitemShut {NoStop}%
\bibitem [{\citenamefont {Rondin}\ \emph {et~al.}(2014)\citenamefont {Rondin},
  \citenamefont {Tetienne}, \citenamefont {Hingant}, \citenamefont {Roch},
  \citenamefont {Maletinsky},\ and\ \citenamefont
  {Jacques}}]{rondinMagnetometryNitrogenvacancyDefects2014}%
  \BibitemOpen
\bibfield  {journal} {  }\bibfield  {author} {\bibinfo {author} {\bibfnamefont
  {L.}~\bibnamefont {Rondin}}, \bibinfo {author} {\bibfnamefont {J.-P.}\
  \bibnamefont {Tetienne}}, \bibinfo {author} {\bibfnamefont {T.}~\bibnamefont
  {Hingant}}, \bibinfo {author} {\bibfnamefont {J.-F.}\ \bibnamefont {Roch}},
  \bibinfo {author} {\bibfnamefont {P.}~\bibnamefont {Maletinsky}},\ and\
  \bibinfo {author} {\bibfnamefont {V.}~\bibnamefont {Jacques}},\ }\href
  {https://doi.org/10.1088/0034-4885/77/5/056503} {\bibfield  {journal}
  {\bibinfo  {journal} {Rep. Prog. Phys.}\ }\textbf {\bibinfo {volume} {77}},\
  \bibinfo {pages} {056503} (\bibinfo {year} {2014})}\BibitemShut {NoStop}%
\bibitem [{\citenamefont {Hingant}\ \emph {et~al.}(2015)\citenamefont
  {Hingant}, \citenamefont {Tetienne}, \citenamefont {Mart{\'i}nez},
  \citenamefont {Garcia}, \citenamefont {Ravelosona}, \citenamefont {Roch},\
  and\ \citenamefont {Jacques}}]{hingantMeasuringMagneticMoment2015}%
  \BibitemOpen
  \bibfield  {author} {\bibinfo {author} {\bibfnamefont {T.}~\bibnamefont
  {Hingant}}, \bibinfo {author} {\bibfnamefont {J.-P.}\ \bibnamefont
  {Tetienne}}, \bibinfo {author} {\bibfnamefont {L.~J.}\ \bibnamefont
  {Mart{\'i}nez}}, \bibinfo {author} {\bibfnamefont {K.}~\bibnamefont
  {Garcia}}, \bibinfo {author} {\bibfnamefont {D.}~\bibnamefont {Ravelosona}},
  \bibinfo {author} {\bibfnamefont {J.-F.}\ \bibnamefont {Roch}},\ and\
  \bibinfo {author} {\bibfnamefont {V.}~\bibnamefont {Jacques}},\ }\href
  {https://doi.org/10.1103/PhysRevApplied.4.014003} {\bibfield  {journal}
  {\bibinfo  {journal} {Phys. Rev. Applied}\ }\textbf {\bibinfo {volume} {4}},\
  \bibinfo {pages} {014003} (\bibinfo {year} {2015})}\BibitemShut {NoStop}%
\bibitem [{\citenamefont {Ramazanoglu}\ \emph
  {et~al.}(2011{\natexlab{b}})\citenamefont {Ramazanoglu}, \citenamefont
  {Ratcliff}, \citenamefont {Choi}, \citenamefont {Lee}, \citenamefont
  {Cheong},\ and\ \citenamefont {Kiryukhin}}]{ramazanogluTemperature2011}%
  \BibitemOpen
  \bibfield  {author} {\bibinfo {author} {\bibfnamefont {M.}~\bibnamefont
  {Ramazanoglu}}, \bibinfo {author} {\bibfnamefont {W.}~\bibnamefont
  {Ratcliff}}, \bibinfo {author} {\bibfnamefont {Y.~J.}\ \bibnamefont {Choi}},
  \bibinfo {author} {\bibfnamefont {S.}~\bibnamefont {Lee}}, \bibinfo {author}
  {\bibfnamefont {S.-W.}\ \bibnamefont {Cheong}},\ and\ \bibinfo {author}
  {\bibfnamefont {V.}~\bibnamefont {Kiryukhin}},\ }\href
  {https://doi.org/10.1103/PhysRevB.83.174434} {\bibfield  {journal} {\bibinfo
  {journal} {Physical Review B}\ }\textbf {\bibinfo {volume} {83}},\ \bibinfo
  {pages} {174434} (\bibinfo {year} {2011}{\natexlab{b}})}\BibitemShut
  {NoStop}%
\bibitem [{\citenamefont {Johnson}\ \emph {et~al.}(2013)\citenamefont
  {Johnson}, \citenamefont {Barone}, \citenamefont {Bombardi}, \citenamefont
  {Bean}, \citenamefont {Picozzi}, \citenamefont {Radaelli}, \citenamefont
  {Oh}, \citenamefont {Cheong},\ and\ \citenamefont
  {Chapon}}]{johnsonXRayImagingMultiferroic2013}%
  \BibitemOpen
  \bibfield  {author} {\bibinfo {author} {\bibfnamefont {R.~D.}\ \bibnamefont
  {Johnson}}, \bibinfo {author} {\bibfnamefont {P.}~\bibnamefont {Barone}},
  \bibinfo {author} {\bibfnamefont {A.}~\bibnamefont {Bombardi}}, \bibinfo
  {author} {\bibfnamefont {R.~J.}\ \bibnamefont {Bean}}, \bibinfo {author}
  {\bibfnamefont {S.}~\bibnamefont {Picozzi}}, \bibinfo {author} {\bibfnamefont
  {P.~G.}\ \bibnamefont {Radaelli}}, \bibinfo {author} {\bibfnamefont {Y.~S.}\
  \bibnamefont {Oh}}, \bibinfo {author} {\bibfnamefont {S.-W.}\ \bibnamefont
  {Cheong}},\ and\ \bibinfo {author} {\bibfnamefont {L.~C.}\ \bibnamefont
  {Chapon}},\ }\href {https://doi.org/10.1103/PhysRevLett.110.217206}
  {\bibfield  {journal} {\bibinfo  {journal} {Phys. Rev. Lett.}\ }\textbf
  {\bibinfo {volume} {110}},\ \bibinfo {pages} {217206} (\bibinfo {year}
  {2013})}\BibitemShut {NoStop}%
\bibitem [{\citenamefont {Waterfield~Price}\ \emph {et~al.}(2016)\citenamefont
  {Waterfield~Price}, \citenamefont {Johnson}, \citenamefont {Saenrang},
  \citenamefont {Maccherozzi}, \citenamefont {Dhesi}, \citenamefont {Bombardi},
  \citenamefont {Chmiel}, \citenamefont {Eom},\ and\ \citenamefont
  {Radaelli}}]{waterfieldpriceCoherentMagnetoelasticDomains2016}%
  \BibitemOpen
  \bibfield  {author} {\bibinfo {author} {\bibfnamefont {N.}~\bibnamefont
  {Waterfield~Price}}, \bibinfo {author} {\bibfnamefont {R.~D.}\ \bibnamefont
  {Johnson}}, \bibinfo {author} {\bibfnamefont {W.}~\bibnamefont {Saenrang}},
  \bibinfo {author} {\bibfnamefont {F.}~\bibnamefont {Maccherozzi}}, \bibinfo
  {author} {\bibfnamefont {S.~S.}\ \bibnamefont {Dhesi}}, \bibinfo {author}
  {\bibfnamefont {A.}~\bibnamefont {Bombardi}}, \bibinfo {author}
  {\bibfnamefont {F.~P.}\ \bibnamefont {Chmiel}}, \bibinfo {author}
  {\bibfnamefont {C.-B.}\ \bibnamefont {Eom}},\ and\ \bibinfo {author}
  {\bibfnamefont {P.~G.}\ \bibnamefont {Radaelli}},\ }\href
  {https://doi.org/10.1103/PhysRevLett.117.177601} {\bibfield  {journal}
  {\bibinfo  {journal} {Phys. Rev. Lett.}\ }\textbf {\bibinfo {volume} {117}},\
  \bibinfo {pages} {177601} (\bibinfo {year} {2016})}\BibitemShut {NoStop}%
\bibitem [{\citenamefont {Xu}\ \emph {et~al.}(2018)\citenamefont {Xu},
  \citenamefont {Dup{\'e}}, \citenamefont {Xu}, \citenamefont {Xiang},\ and\
  \citenamefont {Bellaiche}}]{xuRevisitingSpinCycloids2018}%
  \BibitemOpen
  \bibfield  {author} {\bibinfo {author} {\bibfnamefont {B.}~\bibnamefont
  {Xu}}, \bibinfo {author} {\bibfnamefont {B.}~\bibnamefont {Dup{\'e}}},
  \bibinfo {author} {\bibfnamefont {C.}~\bibnamefont {Xu}}, \bibinfo {author}
  {\bibfnamefont {H.}~\bibnamefont {Xiang}},\ and\ \bibinfo {author}
  {\bibfnamefont {L.}~\bibnamefont {Bellaiche}},\ }\href
  {https://doi.org/10.1103/PhysRevB.98.184420} {\bibfield  {journal} {\bibinfo
  {journal} {Physical Review B}\ }\textbf {\bibinfo {volume} {98}},\ \bibinfo
  {pages} {184420} (\bibinfo {year} {2018})}\BibitemShut {NoStop}%
\bibitem [{\citenamefont
  {Schönherr}(2018)}]{schonherrTopologicalStructuresMagnetic2018}%
  \BibitemOpen
  \bibfield  {author} {\bibinfo {author} {\bibfnamefont {P.}~\bibnamefont
  {Schönherr}},\ }\emph {\bibinfo {title} {Topological {Structures} in
  {Magnetic} and {Electric} {Materials}}},\ \href
  {https://doi.org/10.3929/ethz-b-000269597} {Ph.D. thesis},\ \bibinfo
  {school} {ETH Zurich} (\bibinfo {year} {2018})\BibitemShut {NoStop}%
\end{thebibliography}

%

\end{document}